*energies* MDPI

Article# A critical review of data-driven transient stability assessment of power systems: principles, prospects and challenges

Shitu Zhang [1], Zhixun Zhu [2], Yang Li [1,*]

[1] School of Electrical Engineering, Northeast Electric Power University, Jilin 132012, China;
[2] GHN Energy Jilin Jiangnan Thermal Power CO., LTD, Jilin 132013, China.
* Correspondence: liyang@neepu.edu.cn.**Abstract:** Transient stability assessment (TSA) has always been a fundamental means for ensuring the secure and stable operation of power systems. Due to the integration of new elements such as power electronics, electric vehicles and renewable power generations, dynamic characteristics of power systems are becoming more and more complex, which makes TSA an increasingly urgent task. Since traditional time-domain simulations and direct method cannot meet the actual operation requirements of power systems, data-driven TSA has attracted growing attention from both academia and industry. This paper makes a comprehensive review from the following four aspects: feature extraction and selection, model construction, online learning and rule extraction; and then, summarizes the challenges and prospects for future research; finally, draws the conclusions of this review. This review will be beneficial for relevant researchers to better understand the research status, key technologies and existing challenges in the field.

**Keywords:** transient stability assessment; power systems; data-driven approach; feature extraction and selection; model construction; review## 1. Introduction

Transient stability assessment (TSA) is a fundamental means for ensuring the secure and stable operation of power systems. Transient stability of power systems refers to the ability of each generator in the system to maintain synchronous operation after a large disturbance [1]. With the increasing penetration of new elements such as power electronics, electric vehicles and renewable power generations, dynamic characteristics of power systems are becoming more and more complex. In this situation, accurate and rapid TSA is increasingly urgent. With the rapid development of artificial intelligence techniques, data-driven TSA approaches have become a hot topic in recent years, and a large number of research results have been produced. Therefore, it is necessary to make a critical review of existing data-driven TSA approaches, so that relevant researchers can better understand the research status, key technologies and existing challenges in the field.

As summarized in Table 1, existing TSA methods can be roughly divided into three categories: time-domain simulation method [2], direct method [3], and data-driven artificial intelligence (AI) method [4, 5]. The basic idea of time-domain simulation methods is to use a numerical integration algorithm to solve the differential-algebraic equations (DAEs) describing the dynamic process of a disturbed power system, and then judge the stability status of the system by the relative angle changes between generator rotors. Due to good model adaptability and reliability, this method has been widely used in the electric power industry. Reference [6] proposes the application of the unsymmetric multifrontal method to solve the DAEs encountered in the power system dynamic simulations. Reference [7] proposes a time-domain simulation approach for power system dynamic simulations by using unsymmetric multifrontal method. Reference [8] proposes a distributed transient stability simulation algorithm, which has a good strong scalability. Using the above mentioned technologies, the existing transient simulation can realize super real-time simulations for large-scale power systems.

The direct method is a kind of TSA method that uses energy functions constructed by Lyapunov theory [9] to analyze the transient stability of a power system. Reference [10] reveals the role of the Koopman model in power system transient stability assessment. Compared with the time-domain simulation method, this algorithm does not require complex time-domain simulation of the system after a fault, and it can provide a measure of the degree of system stability. References [11,12] propose a single machine equivalent (SIME) method for transient stability assessment. Reference [13] proposes a method for transient stability assessment of a multi-machine system by using the extended equal area criterion (EEAC). In addition, phasor measurement units (PMUs) [14] and dynamic state estimator (DSE) [15] can collect online information in real time for TSA.

Unlike the above-mentioned time-domain simulation method and direct method, a data-driven TSA method is model-free, which treats TSA as a pattern recognition problem. In this method, an AI-based assessment model is built to reflect the input power system operational parameters and the transient stability status of the system. This method has the advantages of strong learning ability and fast evaluation speed, which has a good performance in the field of power system transient stability assessment.

In order to facilitate analysis, the principles, advantages and disadvantages of different kinds of TSA methods are shown in Table 1.

Table 1. Principles, advantages and disadvantages of different TSA methods

| Methods | Principles | Advantages | Disadvantages |
|---|---|---|---|
| Time-domain simulation | Solve differential-algebraic equations describing the dynamic process of a disturbed power system | This method has good scalability with accurate and reliable results. | The calculation results depend on the accuracy of the system model and parameters. |
| Direct method | Construct an energy function to describe the transient stability of a power system | This method has fast calculation speed and can provide a stability margin. | The energy function is difficult to construct, and the calculation result is conservative. |
| Data-driven TSA | Judge the stability status of a disturbed system using a trained TSA model | The method has strong learning ability and fast calculation speed. | It acts as a black box with poor interpretability and weak adaptability to topological changes |

## 2. Principles of data-driven transient stability assessment

As a mode-free method, data-driven TSA regards transient stability assessment as a pattern classification problem, which mainly includes the following aspects: feature extraction and selection, model construction, online learning and rule extraction. For ease of description, a schematic diagram of data-driven TSA is shown in Fig. 1.

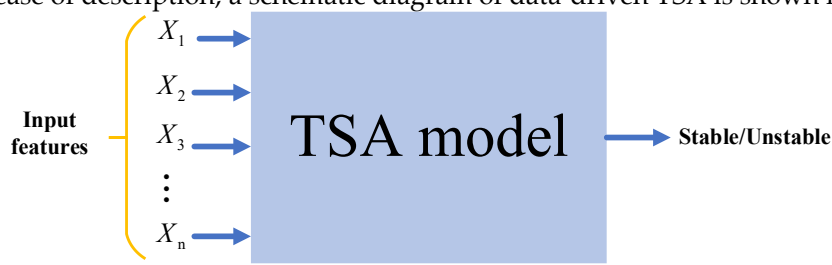

Fig. 1. Schematic diagram of data-driven TSA

In Fig. 1, $X_i$ ($i=1,…,n$) denotes input feature $i$, and all input features constitute a feature vector as the input of TSA models. A TSA model can learn the mapping relationships between input features and system stability status. After an assessment model is trained, once a new input feature vector is sent to the TSA model, the stability status of the system will be immediately predicted by using the mapping relationship obtained through model training.

*2.1. Feature extraction and selection*

Feature extraction and feature selection are two important issues for TSA of power systems. Feature extraction is to extract input features from initial set of measured data, while feature selection refers to the process of choosing a subset of relevant features for use in model construction. Especially, feature selection is a typical combination optimization problem. Compared with traditional machine learning algorithms, deep learning can automatically process all features, generate more complex combined features, and eliminate possible omissions in feature extraction algorithms and the subjective factors of researchers.

Based on the structural risk minimization principle, support vector machines (SVMs) can achieve accurate classification in a small sample space. Reference [16] adopts feature selection methods to determine the input variables best suitable for training an ANN-based TSA model. Reference [17] presents a SVM-based two-stage feature selection method: in the first stage, the original feature set is sorted by the SVM recursive feature selection method, and the unimportant features are eliminated to obtain a reduced feature subset; in the second stage, the SVM with radial basis function kernel is used as the classifier to select approximate optimal feature subsets. Reference [18] proposes a TSA method based on enhanced feature selection and least square SVM. Considering the post-fault measurement information provided by PMUs, reference [19] proposes a feature selection method based on the improved maximal-relevance and minimal redundancy criterion (mRMR) and SVM for transient stability assessment. In reference [20], the feature selection algorithm based on random forest and recursive feature elimination is used to extract the key feature subset for TSA.

With the rapid development of AI technology, deep learning has been successfully applied to the feature extraction and selection of power system transient stability assessment in recent years. Reference [21] proposes a Fisher linear discriminant function method combined with feature selection technology. Fisher discriminator is used to determine the goodness score of each feature, and then rank the features according to their scores. Reference [22] proposes a temporal feature selection method for a time adaptive TSA method, which can extract the crucial temporal features by calculating the feature importance. In reference [23], enhanced feature selection and extraction methods are developed for reducing input features to a probabilistic neural network based TSA model.

In addition, there are other previous works that develop different feature selection methods for data-driven TSA. Reference [24] proposes an mRMR-based mutual information criterion for feature selection. A TSA approach based on the ensemble of OS-extreme learning machine (EOS-ELM) is put forward by using the binary Jaya algorithm to select the optimal feature subset [25]. Reference [26] presents an artificial neural network (ANN) based TSA approach and points out that proper feature selections make this approach a candidate for addressing a topologically independent assessment process.

*2.2. Model construction*

It's known that for general pattern classification problems, constructing an appropriate assessment model is the key to ensuring a proper balance between complexity and generalization, which can avoid the problem of under-learning or over-learning and improve the model's classification performance. Accordingly, model construction is a critical issue for data-driven TSA methods since an appropriate classifier design plays an important role in the performance of the used method.

Existing model construction methods of data-driven TSA mainly include the following categories: ANN, SVM, ensemble learning (EL), and deep learning (DL). These four types of TSA model construction methods are summarized in Table 2.

Table 2. Comparison of different TSA model construction methods

| Categories | Algorithms | Features | Introduction | Reference |
|---|---|---|---|---|
| ANN | Long short-term memory network (LSTM) | Voltage phasor and maximum angle deviation | It proposes a temporal self-adaptive scheme, it aims to balance the trade-off between assessment accuracy and response time. | [27] |
| | Spatial-temporal graph convolutional | Voltage magnitude, active power injection, and reactive power in- | It utilizes graph convolution to integrate network topology information and adopts | [28] |
| | network | jection time series | one-dimensional convolution to exploit temporal information. | |
| | Convolutional neural network (CNN) | Bus voltage | It can not only assess whether the system will be stable or unstable, but also predict the instability mode for the unstable status. | [29] |
| | Recurrent graph convolution neural (RGCN) | Bus voltage magnitude, the bus relative phase and the rotor speeds of generators | It aggregates both the GCN and the LSTM unit to form the RGCN. | [30] |
| SVM | SVM | Generator rotor angles, generator speeds, voltage magnitudes | It can be early predicted based on the measured post-fault values of the generator voltages, speeds, or rotor angles. | [31] |
| | Aggressive SVM (ASVM) and conservative SVM (CSVM) | Active power, reactive power, phase angle of bus voltage, generator information | It proposes a strategy combining grey region and two SVMs to deal with the problems of false alarms and false dismissals. | [32] |
| | Core vector machine (CVM) | Load condition, rotor angle, speed and acceleration | It builds a TSA model based on core vector machine. | [33] |
| | Multi-layer SVM (MLSVM) | Reactive and active power of generators, bus voltage and angle, Reactive and active power of reload | It uses genetic algorithm for a MLSVM-based TSA model to identify valued feature subsets with varying numbers of features. | [34] |
| Ensemble learning | A denoising stacked autoencoder and a voting ensembler | Frequency | It uses cross-entropy to evaluate the fitting performance of base learners and to set the weight coefficient in the ensembler. | [35] |
| | Variational Bayes multiple kernel learning | Voltage/current phasor, active and reactive power, power factor and system frequency | It uses the post disturbance PMU data to predict the system and calculate the stability margin for a given emergency. | [36] |
| | Mahalanobis Kernel | Network topology | It makes efficient use of data under different network topologies, and thus enhances the estimation accuracy and reduces the need for training samples. | [37] |
| | Adaptive ensemble decision tree (DT) | Voltage magnitudes, active/reactive power flows and current flows, voltage phase angle differences | It proposes an adaptive ensemble DT learning based TSA approach considering operating condition variations and topology changes. | [38] |
| Deep learning | Deep belief network | Steady-state features, transient features, fault removal features | It initializes with unsupervised learning using unlabeled samples, and then fine-tune with supervised learning using labeled samples. | [40] |

| | | | |
|---|---|---|---|
| Stacked autoencoder (SAE) | Static features, system-level classification features, system-level classification features, | It proposes a SAE based feature reduction method for TSA. | [41] |
| CNN and LSTM | Voltage phasor measurements | It presents a unified deep learning prediction model for small signal and transient stability. | [42] |

*2.2.1. ANN-based TSA*

ANN is a widely-used AI algorithm for addressing transient stability assessment problems. In such kind of data-driven TSA method, ANN is utilized to build a TSA model reflecting the mapping relationships between power system operational parameters and the system stability status.

In reference [26], artificial neural network is used to construct a TSA model for the first time [26]. Reference [27] develops a temporal self-adaptive TSA system by using long short-term memory network (LSTM), which can learn the time dependence of the input temporal sequences. A spatial-temporal graph convolutional network is put forward for TSA of power systems in reference [28]. Reference [29] proposes a TSA and instability mode prediction model based on convolutional neural network (CNN). A multi-task TSA framework is proposed for power systems by using recurrent graph convolutional networks (RGCN) [30].

*2.2.2. SVM-based TSA*

Compared with traditional ANNs, SVM has better generalization ability and stability. Reference [31] introduces a SVM based TSA model and compares it with the common multilayer perceptron models. A real-time TSA approach is presented for power system based on improved SVMs in reference [32]. The improved SVMs include aggressive support vector machine (ASVM) and conservative support vector machine (CSVM). By using big data and the core vector machine (CVM), a TSA method is established in reference [33]. A multi-layer SVM (MLSVM) optimized by genetic algorithm (GA) is put forward for transient stability assessment of power systems in reference [34]. The results show that the MLSVM is able to reduce the possibility of misclassification of transient stability assessment.

*2.2.3. Ensemble learning-based TSA*

The key idea of ensemble learning is to combine multiple learners into an algorithm model with stronger generalization performance by combining strategies. In order to analyze transient stability problems, a complete machine learning-based TSA model is proposed for TSA by using a denoising stacked autoencoder and a voting ensemble classifier [35]. A variational Bayes multiple kernel learning (VBpMKL)-based TSA model is built using multi-feature fusion through combining feature spaces corresponding to each feature subset, it can improve the accuracy and reliability of classification [36]. Reference [37] proposes a data-driven TSA model based on Mahalanobis kernel regression and ensemble learning taking into account network topology changes. In reference [38], an adaptive ensemble learning model based on decision tree (DT) is established to adapt to the changes of system operating conditions and line topology in dynamic security assessment. By this means, the ensemble learning solves the problem of accuracy fluctuations of a single prediction model and greatly improves the reliability of the evaluation results.

*2.2.4. Deep learning-based TSA*

Due to the powerful feature learning and data mining capabilities, deep learning has been widely used to build power system stability assessment models in recent years [39]. In reference [40], a TSA method based on deep belief networks is proposed, and test results show that the presented method performs very well with insufficient training samples or redundant features. Reference [41] puts forward a transient stability assessment method based on deep learning, which constructs three parts of the original feature set and utilizes the stacked autoencoder (SAE) to extract multi-level features. Test results verify that the presented approach is able to reduce the training burden of the assessment model and improve the model's accuracy. Reference [42] proposes a unified deep learning prediction model for analyzing small signal and transient stability of power systems. It uses a CNN-based classifier to determine the transient stability of the system, and then adopts LSTM network to capture low-frequency oscillatory response of a predicted stable system.

*2.3. Online learning*

Since that power system is a time-varying system and training samples generated by offline simulations cannot cover all the operating conditions of the system, the TSA model obtained through offline training may not have good applicability in practical applications. For this reason, it is of great significance to study the online learning ability of TSA models.

Research shows that online support vector regression (SVR) is an effective online learning algorithm for super short-term load forecasting of power systems [43]. Furthermore, reference [44] presents a comprehensive transient stability classifier based on improved SVM, which can speed up the training speed by decomposing large-scale training into parallel small-scale training. Reference [45] proposes a TSA method based on the online sequential extreme learning machine (OS-ELM), which can update the assessment model on-line by partial training. A hierarchical deep learning machine (HDLM) based TSA model is presented to achieve quantitative and qualitative stability prediction in reference [46]. Reference [47] presents a TSA method based on dual cost-sensitivity factors, which can achieve online updating of the model by using incremental learning. In reference [25], an EOS-ELM based TSA model is presented to implement online model updating by using OS-ELM as a weak classifier and the online boosting algorithm as ensemble learning algorithm. In future work, it can use the proposed model as a trigger for wide-area protection. However, it does not take into account the possible PMU failures and communication delays of wide area measurement systems (WAMS) that may occur in real-world power systems, which is shown in Fig. 2.

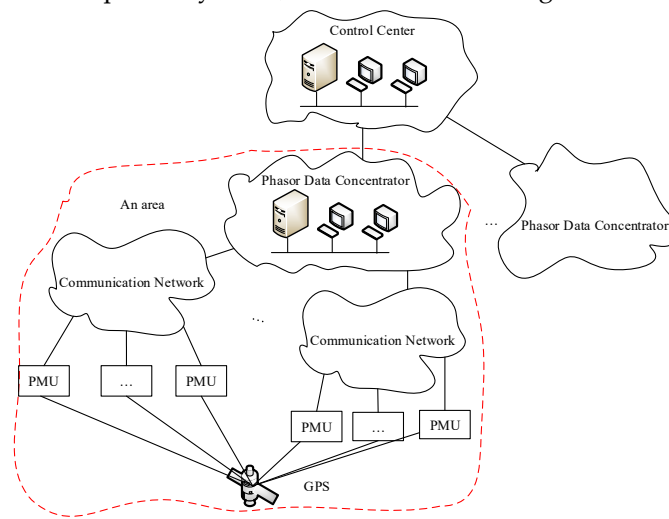

Fig. 2. Structure of WAMS

As shown in Fig. 2, WAMS is a measurement system for a power system including multiple areas, and it typically consists of three components: PMUs, communication system and control system.

*2.4. Rule extraction*

The traditional research on transient problems starts from the physical mechanism, and mainly includes the numerical integration method based on mathematical modeling and the direct method of analyzing the energy conversion of the system. Different from time domain simulation or energy function methods, data-driven TSA approaches regard the power system as a "black box" system to fit the relationship between input and output. Rule extraction is an important problem of the "black box" machine learning system, and its purpose is to express the knowledge learned in the learning machine in an easy-to-understand way.

There have already been some previous works in the field of TSA rule extraction. For example, a method for extracting transient stability rules based on linear decision trees is proposed in reference [48], which screens the support samples near the stable boundary as the input samples of the decision tree and reduces the number of samples to obtain safe and stable operation rules based on combined attributes. However, the evaluation results are sensitive to sample composition and the extension ability and robustness of decision-making knowledge are poor. Furthermore, reference [49] proposes a method for extracting power grid transient stability rules based on multi-attribute decision trees. A decision tree-based TSA model is constructed after the discretization of the transient stability margins under some specified faults, then the general rule for evaluating the stability of the system is achieved. However, this refer-ence does not consider economic factors. A method for extracting transient stability assessment rules based on extreme learning machine (ELM) and improved ant-miner (IAM) algorithm is proposed in reference [50], which has important research value for improving the comprehensibility and interpretability of TSA methods. However, it is necessary to ensure that the generated samples fully reflect the response characteristics of the training model and cover the entire sample space with a uniform distribution. Reference [51] proposes a power system stability assessment and rule extraction approach based on pattern discovery. However, this method only analyzes a single fault and does not extend to cascading faults.

*2.5. Overall flowchart of data-driven TSA*

For ease of presentation, an overall flowchart of a typical data-driven TSA approach is shown in Fig. 3.

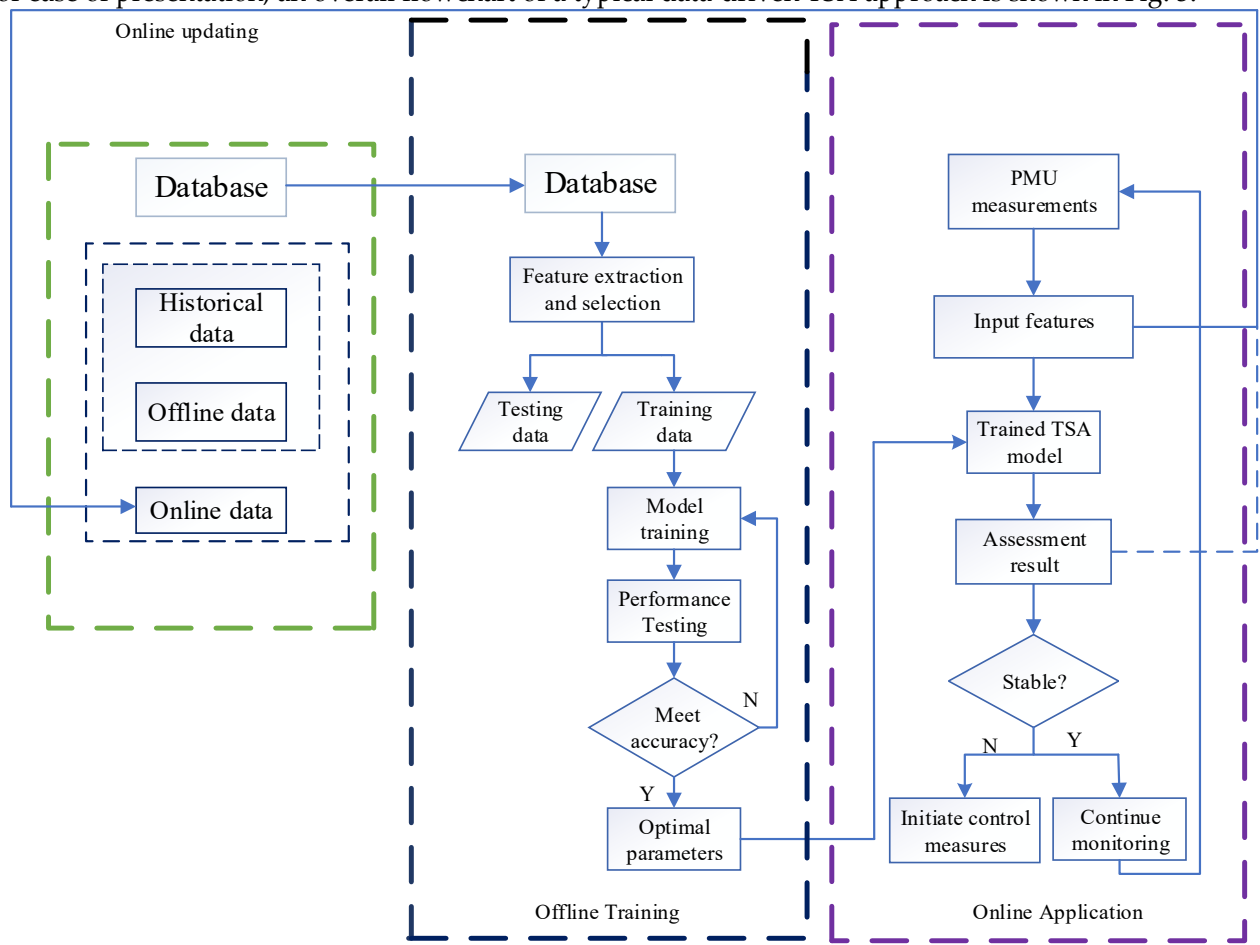

**Fig. 3. Flowchart of a typical data-driven transient stability assessment method.**

As shown in Fig. 3, the overall process of a typical data-driven TSA can be roughly divided into two stages: offline training and online application. In the stage of offline training, feature extraction and selection are performed to provide appropriate input features, and then an assessment model is constructed to find the right balance between complexity and generalization during the classification process. The assessment model will be trained continuously until its performance meets the expected requirement. In the stage of online application, once an input feature vector obtained from the measured data reaches a trained TSA model, the stability status will be predicted. If the predicted result is unstable, control measures will be initiated at once; otherwise, the evaluation process will continue into the next monitoring cycle. Note that, besides offline data, historical archives of power system operation and new samples generated online can be incorporated into training databases to train/update the assessment model.

**3. Future challenges and prospects**

*3.1. Impact of renewable energy integration*

At present, today's power system is in a transitional stage toward a hybrid power system with high penetration of renewables. In recent years, the proportion of renewable energy resources in power systems has been increasing [52]. As a result, the dynamic characteristics of power systems are becoming more and more complex. As an important form of renewable energy integration, microgrids have been vigorously promoted in power systems [53]. Taking into account the uncertainties of load and renewable power generation, a chance constrained programming based optimal dispatch model of isolated microgrids with energy storage is proposed in reference [54]. However, as a kind of clean energy, renewable energy has inherent uncertainties [55, 56], which will affect the secure operation of power systems. With the grid connection of large wind farms, the power flow of the power system will also change and interact with synchronous generators to affect the small signal angle stability of the power system [57]. Reference [58] considers the impact of different types and capacities of wind power generators on the power grid, and then investigates the influence of large-scale renewable integration on the transient stability of a power system. Reference [59] studies the equivalent modelling of hybrid renewable energy source plants for TSA. A Probabilistic TSA method is proposed for power systems with renewables in reference [60].

The above comprehensive analysis shows that power systems are becoming increasingly complex due to the integration of renewables. Therefore, the stability analysis of power grid operation has gradually shifted to online analysis, which is an inevitable result of the development of power grid systems. However, there are still many challenges and difficulties in the analysis process. Therefore, the development process of analysis methods is relatively slow. The relevant theories for system large interference problem analysis are mature, and there are relatively many types of software that can be used for calculation. However, the application of software and the limited control of stable operation of the power grid is relatively difficult, and in-depth exploration by relevant industry personnel is required.

*3.2. Stability assessment of AC/DC systems with VSC-HVDC*

Voltage source converter based high voltage direct current transmission (VSC-HVDC) is a new type of direct current transmission technology. Wind farms are connected to the grid through the VSC-HVDC system, and the voltage stability, power quality, and penetration power can be significantly improved. Especially in some areas where the development of wind power is encountering difficulties, the use of VSC-HVDC provides an effective way to address the technical problems of long-distance and large-capacity wind power transmission. However, when the voltage drops, the active power sent by a high voltage direct current (HVDC) converter station is greatly reduced, which will cause the power imbalance between the receiving and sending ends of the converter station [61]. In some severe cases, the devices will be damaged or even the HVDC lines will be tripped, leading to the system instability and the failure of low voltage ride-through (LVRT). Therefore, how to assess the stability of AC/DC systems with VSC-HVDC has become a hot topic in the field of electrical engineering.

As a type of direct current transmission technology, voltage source converter based high voltage direct current transmission (VSC-HVDC) has received extensive attention from researchers due to its flexible and fast adjustment capabilities. Reference [62] proposes a two-stage solution method combining multi-objective optimization and decision support by using the non-dominated sorting genetic algorithm II (NSGA-II). Reference [63] proposes a two-stage AC/DC system multi-objective optimal power flow (MOPF) method that integrates decision analysis into the optimization process. Reference [64] presents a controlled islanding model for AC/DC systems with VSC-HVDC to minimize the source-load imbalance by using semi-supervised spectral clustering, which uses VSC-HVDC links for power exchanges between islands. Reference [65] puts forward a black-start strategy based on VSC-HVDC for passive networks and carries out a number of simulation tests. The test results shows that VSC-HVDC can improve the system stability during the recovery process and shorten the system recovery time.

*3.3 Stability assessment considering network topology changes*

Deep neural network have completely changed machine learning tasks. Although convolution neural network is widely used, they have limitations in processing non-Euclidean spatial data. Graph neural network plays an important role in the application of non-Euclidean data in deep learning, especially the use of graph structures that can be explained on traditional Bayesian causal networks. It is of great significance to define the inferable and causal interpretable problems of deep neural network relationships. Therefore, how to use deep learning technology to analyze and reason about graph structure data has attracted widespread attention from scholars.

A large number of existing studies have shown that transient instability of power systems exhibits a certain spatial distribution characteristics. An energy margin analytical sensitivity method is developed for TSA of power systems considering topology changes in reference [66]. A TSA method based on ANN is presented with consideration of topology changes in reference [67]. Reference [68] has verified the influence of topology on power system transient and frequency stability, by studying the four network topologies: random graph, small-world graph, nested small-world graph, and lattice graph. A TSA approach based on ensemble learning and kernel regression has been presented taking into account topology changes in reference [69]. Based on a message transfer graph neural network, a fast transient stability assessment method based on steady-state data is proposed for power systems in reference [70]. In this method, a TSA model that can describe power system topology changes can be trained by using graph data processing and topological modeling.

*3.4 Limitations in applications and prospects*

Currently, data-driven transient stability assessment methods face some challenges. First, it is tough and expensive to obtain large-scale, balanced data with accurate labels in real-world applications [71, 72]. Then, existing data-driven TSA methods act as a black box with poor interpretability [73, 74], which also limits their application in actual power systems. Finally, most of data-driven TSA methods generally lack the adaptability to topological changes.

At the same time, some emerging techniques are beneficial for developing advanced data-driven TSA methods. Firstly, data augmentation based deep generative learning is a promising technique for addressing complex data analysis issues such as class imbalance and missing data [75, 76]. Secondly, cutting-edge artificial intelligence techniques are helpful to build a powerful TSA model. For example, use of automated reinforcement learning is able to automatically determine the optimal model parameters of an assessment model [77]. It's also an interesting topic to balance accuracy and response speed by using multi-objective optimization [78, 79]. Thirdly, the rapid improvement of software and hardware technology provides powerful computing power for data-driven TSA methods [80-82].

**4. Conclusions**

With the integration of power electronic equipment and renewable energy resources, today's power systems are evolving towards a new generation of power systems with high-penetration renewable energy and power electronics. These changes pose huge challenges for transient stability assessment of power systems. Unlike traditional time domain simulation and energy function methods, data-driven TSA methods establish the relationship between the system operational parameters and the stability status and then directly determine the stability results, which does not require the physical model and parameter information of a power system.

Fast and accurate transient stability assessment plays a crucial role in ensuring the secure and stable operation of power systems. This review article summarizes data-driven transient stability assessment methods from four aspects, i.e., feature extraction and selection, model construction, online learning, and rule extraction. And then, it discusses main challenges and the future development direction in the field. This review will be helpful for relevant researchers to better understand the research status, key technologies and existing challenges in the area of data-driven transient stability assessment of power systems.

**Author Contributions:** The paper was a collaborative effort among the authors. All authors have read and agreed to the published version of the manuscript.

**Acknowledgments:** This work is partly supported by the Natural Science Foundation of Jilin Province, China under Grant No. YDZJ202101ZYTS149.

**Conflicts of Interest:** The authors declare no conflict of interest.